 \documentclass[11pt]{article}
 \usepackage[top=1in,bottom=1in,left=1.5in,right=1.5in]{geometry}
  \usepackage{amsmath,amssymb}
  \usepackage{latexsym}
\usepackage{graphicx,color} 

  \newcommand{\C}{\mathbb{C}}
  \newcommand{\F}{\mathbb{F}}
  \newcommand{\N}{\mathbb{N}}
  
  \newcommand{\R}{\mathbb{R}}

\newcommand{\ii}{\mathbf{i}}

  \newcommand{\e}{\mathbf{e}}
  \newcommand{\f}{\mathbf{f}}

  \newcommand{\h}{\mathbf{h}}
  \newcommand{\g}{\mathbf{g}}

  \newcommand{\U}{\mathbf{U}}
  \newcommand{\uu}{\mathbf{u}}
  \newcommand{\vv}{\mathbf{v}}
  
  \newcommand{\w}{\mathbf{w}}
  
  \newcommand{\x}{\mathbf{x}}
  \newcommand{\X}{\mathbf{X}}
  \newcommand{\y}{\mathbf{y}}
  
  \newcommand{\z}{\mathbf{z}}
  \newcommand{\0}{\mathbf{0}}

  \newcommand{\cC}{\mathcal{C}}
  \newcommand{\cD}{\mathcal{D}}

  \newcommand{\cL}{\mathcal{L}}
  \newcommand{\cM}{\mathcal{M}}
  
  \newcommand{\cP}{\mathcal{P}}
  \newcommand{\cR}{\mathcal{R}}
  
  \newcommand{\cT}{\mathcal{T}}

  \newcommand{\cW}{\mathcal{W}}
  
  \newcommand{\cY}{\mathcal{Y}}

  \newcommand{\rA}{\mathrm{A}}

  \newcommand{\rH}{\mathrm{H}}

  \newcommand{\lan}{\langle}
  \newcommand{\ran}{\rangle}
  \newcommand{\an}[1]{\lan#1\ran}
  \def\diag{\mathop{{\rm diag}}\nolimits}
  \newcommand{\hs}{\hspace*{\parindent}}
  \newcommand{\proof}{\hs \textbf{Proof.\ }}
  
  \newcommand{\tr}{\mathop{\mathrm{tr}}\nolimits}

  \newcommand{\trans}{^\top}
  \newcommand{\qed}{\hspace*{\fill} $\Box$\\}

  \newcommand{\rS}{\mathrm{S}}

  \newcommand{\rank}{\mathrm{rank\;}}

  \newtheorem{theo}{\bfseries \hs Theorem}
  \newtheorem{defn}[theo]{\bfseries \hs Definition}

  \newtheorem{lemma}[theo]{\bfseries \hs Lemma}
  \newtheorem{corol}[theo]{\bfseries \hs Corollary}

 \numberwithin{equation}{section} 

 \renewcommand{\span}{\mathrm{span}}

\begin{document}

 \title{On the extreme points of quantum channels}
\author{{ Shmuel  Friedland\footnote{Supported by NSF grant DMS--1216393.} and Raphael Loewy}\\ \\
Department of Mathematics, Statistics, and Computer Science,\\
        University of Illinois at Chicago\\
        Chicago, Illinois 60607-7045, USA\\
         \and
Department of Mathematics\\
Technion -- Israel Institute of Technology\\
32000 Haifa, Israel}

 \renewcommand{\thefootnote}{\arabic{footnote}}
 \date{October 28, 2014}
 \maketitle
 \begin{abstract}
 Let $\cL_{m,n}$ denote the convex set of completely positive trace preserving operators from $\C^{m\times m}$ to $\C^{n\times n}$, i.e
 quantum channels.  We give a necessary condition for $L\in\cL_{m,n}$ to be an extreme point.  We show that generically, this condition is also sufficient.
 We characterize completely the extreme points of $\cL_{2,2}$ and $\cL_{3,2}$, i.e. quantum channels from qubits to qubits and from
qutrits to qubits.

 \end{abstract}



 \renewcommand{\thefootnote}{\arabic{footnote}}

 \section{Introduction}\label{sec:intro}
 Denote by $\C^{m\times n},\rH_m\supset \rH_{m,+}\supset\rH_{m,+,1}$ the space of complex $m\times n$ matrices,
 $m\times m$ hermitian matrices, the cone of nonnegative definite matrices and the set of density matrices respectively.
 For $A\in\C^{m\times m}$ we denote $A\ge 0$ if and if $A\in\rH_{m,+}$.
 Denote by $\cP_m\subset \rH_{m,+,1}$
the set of all pure states in $\rH_{m,+,1}$, i.e. all rank one hermitian matrices of order $m$ with trace one.
 Let $[m]=\{1,\ldots,m\}$ for any positive integer $m$.
 Recall that $L:\C^{m\times m}\to \C^{n\times n}$ is called completely positive if
 \begin{equation}\label{defcp}
 L(X)=\sum_{i=1}^k A_i XA_i^*, \quad A_i\in\C^{n\times m}, i\in[k].
 \end{equation}
 Observe that
 \begin{equation}\label{trivdefLX}
L(X)=\sum_{i=1}^k B_iXB_i^*, \quad B_i=\zeta_iA_i,\;|\zeta_i|=1,\;i\in[k].
\end{equation}
 $L$ is called quantum channel if $L$ is completely positive and $L:\rH_{m,+,1} \to \rH_{n,+,1}$.  This is equivalent to the statement that
 \begin{equation}\label{trprescon}
 \sum_{i=1}^k A_i^*A_i= I_m,
 \end{equation}
 i.e. $L$ is trace preserving: $\tr L(X)=\tr X$ for all $X\in\C^{m\times m}$.
 Denote by $\cL_{m,n}$ the convex set of all quantum channels $L:\rH_{m,+,1}\to\rH_{n,+,1}$.
 The aim of this note is to study the extreme points of $\cL_{m,n}$.
 We reprove and extend some of the results in \cite{Cho75, RSW, Rus07}.
 Some related results are discussed in \cite{BPS, Tsu93, Tsu96, FM97}.  One of the novel features of this paper is the use of the notions and results
of complex and real algebraic geometry, and semi-algebraic geometry.

The paper is organized as follows. In Section 2 we state Choi's theorem, characterizing a completely positive operator
$L:\C^{m\times m}\to \C^{n\times n}$ in terms of a suitable matrix representation $Z(L)$ in $\C^{mn\times mn}$, and point
out a relation between $\rank Z(L)$ and the number of summands in the representation \eqref{defcp} of $L$. In Section 3
extreme points of the compact convex set $\cL_{m,n}$ are considered.
We give a necessary and sufficient condition for $L$ to be an extreme point of $\cL_{m,n}$ in terms of the
null space of the matrix $Z(L)$. For this purpose we give a new proof of the result stating that if $L\in\cL_{m,n}$ is an extreme point then $\rank Z(L)\in [m]$.


Conversely, in Section 4 we show that a generic $L\in\cL_{m,n}$
with $\rank Z(L)=m$ is an extreme point of $\cL_{m,n}$.
Information on $L$ when $\rank Z(L) \le m$ is also given.
In Section 5 we give a dimension condition on $L\in \cL_{m,n}$
so that $L(\rH_{m,+,1})$ contains a density matrix of rank at most $p$.
In Section 6 we fully characterize the extreme points of $\cL_{2,2}$,
namely we show that $L\in\cL_{2,2}$ is an extreme point of that set if and only if either $L$ is a unitary similarity
transformation, or $\rank Z(L)=2$ and $L$ is not a convex combination of two distinct unitary similarity transformations.
This is known, but our approach is new.
In Section 7 we characterize the extreme points of $\cL_{3,2}$.
Section 8 contains a brief discussion of entropy of quantum channels.

 \section{Preliminary results}
Let $\F$ be the field of real or complex numbers $\R,\C$ respectively.

\noindent
Let  $\e_i=(\delta_{1i},\ldots,\delta_{mi})\trans,i\in[m]$ be the standard basis in $\R^m$.  Introduce the following isomorphism
$\phi_m:\C^{mn}\to \C^{n\times m}$.  View $\z\in\C^{mn}$ as the column vector $\z=(\z_1\trans,\ldots,\z_m\trans)\trans$, where
$\z_i\in\C^n, i\in [m]$. Then $\phi_m(\z)=[\z_1\ldots\z_m]\in\C^{n\times m}$ is the matrix whose columns are $\z_1,\ldots,\z_m$.
Vice versa, if $Z=[\z_1\ldots\z_m]\in \C^{n\times m}$ then $\phi_m^{-1}(Z)$, denoted as $\hat Z$, is the vector
$\z=(\z_1\trans,\ldots,\z_m\trans)\trans$.  Recall that the standard inner product in $\C^{mn}$ given as $\an{\uu,\vv}=\vv^*\uu$
corresponds to the inner product $\an{U,V}=\tr V^*U$ in $\C^{n\times m}$, which is preserved under the isomorphism
$\phi_m$.

 Let $\rS_m(\F),\rA_m(\F)\subset \F^{m\times m}$ be the $\F$ subspases of symmetric and
skew-symmetric matrices respectively.  So $\rS_m(\F)\times \rA_m(\F)$ is viewed as a linear space of tuples $(A,T)$, i.e.
\[\rS_m(\F)\times \rA_m(\F)\sim \rS_m(\F)\oplus \rA_m(\F)\sim\F^{m^2}.\]

Recall that $\rH_m$ can be identified with $\rS_m(\R)\times \rA_m(\R)$ as follows.  We view a hermitian matrix $H=S+\ii T$, where
$S=[s_{ij}],T=[t_{ij}]\in\R^{m\times m}$  are real  symmetric and skew symmetric matrices respectively.
 So $\psi: \rH_m\to \rS_m(\R)\times \rA_m(\R)$  is given by $S+\ii T\mapsto (S,T)\in \rS_m(\R)\times \rA_m(\R)$.

We next observe that $\C^{m\times m}$ is the following complexification of $\rH_m$.  Namely $C\in \C^{m\times m}$ is uniquely represented as $S+\ii T$,
where $S\in \rS_m(\C), T\in\rA_m(\C)$.  So we have the linear map $\psi_m:\C^{m\times m}\to \rS_m(\C)\times \rA_m(\C)$ that is given as above. 
The standard inner product on $\C^{m\times m}$ that is
given by $\an{C_1,C_2}=\tr  C_2^* C_1$ induces the standard inner product on $\rS_m(\C)\times \rA_m(\C)$
\[\an{(S_1,T_1),(S_2,T_2)}:=\tr (S_2^*S_1+T_2^* T_1)=\tr (\bar S_2S_1-\bar T_2T_1).\]
This is the standard inner product on $\C^{m^2}$.  Note that on $\rH_m$ and $\rS_m(\R)\times\rA_m(\R)$ this inner product is real valued.

Denote by $\cL(\rH_m,\rH_n)$ the set of all linear transformations from $\rH_m$ to $\rH_n$.
Let $L\in\cL(\rH_m,\rH_n)$.  Then $L$ is represented by a real matrix $M\in \R^{n^2\times m^2}$.
We now consider the adjoint transformation $L^\vee\in\cL(\rH_n,\rH_m)$.  That is, for
\[\an{L(H_1),H_2}=\an{H_1,L^\vee(H_2)} \textrm{ for all } H_1\in\rH_m, H_2\in\rH_n.\]
So $L^\vee$ is represented by $M\trans\in \R^{m^2\times n^2}$.   Clearly, $(L^\vee)^\vee=L$.
Denote by $\cL_{m,n}^\vee$ the convex set of all linear transformations from $\rH_m$ to $\rH_n$ which are completely
positive and unital, that is, send the identity to the identity. It is well known that if $L$ is given by \eqref{defcp}
then $L^\vee$ is given by
\begin{equation}\label{ladj}
 L^\vee(Y)=\sum_{i=1}^k A_i^* YA_i,  \quad Y\in \C^{n\times n}.
\end{equation}

Clearly, $L\in\cL(\rH_m,\rH_n)$ extends to a linear transformation from $\C^{m\times m}$ to $\C^{n\times n}$, which we denote by $L$, and no confusion will arise. Clearly,
\begin{equation}\label{conjLprop}
L(X^*)=L(X)^*.
\end{equation}
Hence, with any  $L\in\cL(\rH_m,\rH_n)$ we can associate the following matrix
\begin{equation}\label{defZL}
 Z(L):=[Z_{ij}(L)]_{i=j=1}^m\in \C^{mn\times mn},  \textrm{ where }Z_{ij}(L)=L(\e_i\e_j\trans)\in\C^{n\times n}, i,j\in [m].
\end{equation}
Clearly, $Z(L)$ is hermitian.  Vice versa, it is straightforward to show that any $ Z=[Z_{ij}]_{i=j=1}^m\in \rH_{mn}$ induces a unique $L\in\cL(\rH_m,\rH_n)$
such that $Z=Z(L)$.  $\rank Z(L)$ is sometimes called \emph{Choi} rank of $L$ \cite{RSW,Rus07}.

The following theorem is well known.  It includes Choi's necessary and sufficient condition for a completely positive operator \cite{Cho75},  which is part \emph{4} of this theorem.
We bring its proof for completeness.
\begin{theo}\label{choithm}
Let $L\in\cL(\rH_m,\rH_n)$ and assume that $Z(L)$ is given by \eqref{defZL}.
Assume that $l=\rank Z(L)$.
\begin{enumerate}
\item
 Let $\lambda_1,\ldots,\lambda_l$ be the $l$ nonzero eigenvalues of $Z(L)$ with the corresponding
orthonormal eigenvectors $\hat B_1,\ldots, \hat B_l$.  Then
\begin{equation}\label{LformZ}
L(X)=\sum_{i=1}^l \lambda_i B_i XB_i^*.
\end{equation}
\item
\begin{equation}\label{LveeformZ}
L^\vee(Y)=\sum_{i=1}^l \lambda_i B_i^* YB_i.
\end{equation}
So $Z(L^\vee)$ has rank $l$, $l$ nonzero eigenvalues $\lambda_1,\ldots,\lambda_k$ with with the corresponding
orthonormal eigenvectors $\widehat{B_1^*},\ldots, \widehat{B_l^*}$.
\item An operator $L:\C^{m\times m}\to \C^{n\times n}$ is completely positive if and only
 if the matrix $Z(L)$ is nonnegative definite.
\item
Assume that $L$ is completely positive and nonzero.  Then for any representation \eqref{defcp} one has inequality $k\ge l(\ge 1)$.
For $k=l$ there exists a representation of $L$ of the form  \eqref{defcp} where $A_1,\ldots,A_k$ are nonzero orthogonal matrices in $\C^{n\times m}$:
\begin{equation}\label{orthcond}
\tr A_j^*A_i=0 \textrm{ for } i\ne j, \quad \tr A_i^* A_i>0 \textrm{ for } i\in [k].
\end{equation}
\item
Vice versa, suppose that $L$ is given by \eqref{defcp} and the conditions \eqref{orthcond} hold.  Then $l=k$ and the set $\hat A_1,\ldots,\hat A_k$
is a set of $k$ orthogonal eigenvectors of $Z(L)$ corresponding to the eigenvalues $\tr A_1^*A_1,\ldots,\tr A_k^*A_k$ respectively.

In particular if $\tr A_i^*A_i\ne \tr A_j^*A_j$ for $i\ne j$ then the representation \eqref{defcp} of $L$, where the conditions \eqref{orthcond}
hold, is unique up to the change \eqref{trivdefLX} and permuting the order in the summation in \eqref{defcp}.

\end{enumerate}
\end{theo}
\proof  For  $A\in\C^{n\times m}\setminus\{0\}$ we define the operator $L_A(X):=AXA^*$.
Clearly, $Z(L_A)=[(A\e_i)(A\e_j)^*]_{i,j\in[m]}=\hat A (\hat A)^*$.    Hence $Z(L_A)$ is hermitian, nonnegative definite,
 rank one and $\tr Z(L_A)=\tr A^*A$.  That is, the nonzero eigenvalue of $Z(L_A)$ is $\tr A^*A$.  Clearly, $Z(aL_A)=a\hat A (\hat A)^*$ for $a\in\R$.
Assume that $a\in\R\setminus\{0\}$.
Then $\rank Z(aL_A)=1$ and the nonzero eigenvalue of $Z(aL_A)$ is $a\tr A^*A$.  Vice versa, if $Z\in\rH_{mn}$ and $\rank Z=1$
then $Z=a\uu\uu*$ for $a\in\R\setminus\{0\}, \uu\in \C^{mn}\setminus\{\0\}$.  Hence $Z=Z(aL_A)$ where $A=\phi_m(\uu)$.

Recall that we defined $l=\rank Z(L)$.  Then spectral decomposition of $Z(L)$ is
\begin{equation}\label{specdecZL}
Z(L)=\sum_{i=1}^l \lambda_i\uu_i\uu_i^*, \quad \uu_j^*\uu_i=\delta_{ij},\quad \lambda_i\in\R\setminus\0\}, \quad i,j\in [l].
\end{equation}
Let $B_i:=\phi_m(\uu_i)$ for $i\in[l]$.  The above arguments yield \eqref{LformZ}.  This establishes \emph{1}.

It is easily seen that \eqref{LveeformZ} follows from \eqref{LformZ}.
As $\lambda_i\ne 0,
\tr B_i B_j^*=\delta_{ij}$ for $i,j\in[l]$ we deduce that $Z(L^\vee)$ has rank $l$, $l$ nonzero eigenvalues $\lambda_1,\ldots,\lambda_l$
and corresponding orthonormal eigenvectors $\widehat{B_1^*},\ldots,\widehat{B_l^*}$.   This establishes \emph{2}.

Suppose that $L=aL_A$ for some $a\in\R\setminus\{0\}, A=[a_{ij}]\in \C^{n\times m}\setminus\{0\}$.  Then $Z_{ij}(aL_A)=a[a_{pi}\bar a_{qj}]_{p,q\in n}$.
Clearly, $L^\vee=aL_{A^*}$.  So $Z_{pq}(L^\vee)=a[\bar a_{pi} a_{qj}]_{i,j\in[m]}$.  This proves \emph{3} for $L=aL_A$.  Use \emph{1} to deduce \emph{3}
for any $L\in\cL(\rH_m,\rH_n)$.

Assume now that $L$ has a representation  \eqref{defcp}, where we assume that each $A_i$ is nonzero.   Then
$Z(L)=\sum_{i=1}^k Z(L_{A_i})$.  Hence $Z(L)$ is hermitian nonnegative definite and $l\le k$.

Suppose now that $Z(L)\in\rH_{mn,+}$.   Consider the spectral decomposition \eqref{specdecZL}.  Since $Z(L)$ is nonnegative definite $\lambda_i>0$ for $i\in[l]$.
So each $\lambda_i\uu_i\uu_i^*$ induces the linear operator $L_{A_i}$ where $A_i=\phi_m(\sqrt{\lambda_i}\uu_i)$.
Hence $L=\sum_{i=1}^l L_{A_i}$.  This establishes \emph{4}.

Recall that $\sqrt{\lambda_i\lambda_j}\uu_j^*\uu_i=\tr A_j^*A_i$.
This establishes \emph{5}.

Assume we have a decomposition  \eqref{defcp} and \eqref{orthcond}, then we have the decomposition  \eqref{specdecZL} where $\uu_i=\hat A_i$ for
$i\in [k]$.  Suppose that $\tr A_i^*A_i\ne\tr A_j^* A_j$ for $i\ne j$.  Then the spectral decomposition \eqref{specdecZL} is unique up to a permutation of summands.
This shows that the representation \eqref{defcp} of $L$, where the conditions \eqref{orthcond}
hold is unique up to the change \eqref{trivdefLX} and the order in the summation in \eqref{defcp}.  This establishes \emph{6}.\qed

 Note that  Choi's theorem characterizes completely positive $L\in \cL(\rH_m,\rH_n)$ by considering
 a hermitian matix $Z(L)$, which has $(mn)^2$ real parameters.  So $Z(L)$ induces a corresponding matrix $M\in \R^{n^2\times m^2}$  which we discussed above.

\section{Extreme points of quantum channels I}
In this section we give a necessary and sufficient condition for $L\in\cL_{m,n}$ to be an extreme point, stated
in terms of the null space of $Z(L)$. The proof depends also on a subspace $\cW$ defined in the next theorem, whose proof
follows from Theorem \ref{choithm}.


\begin{theo}\label{charqc}  Let
\begin{eqnarray}\label{quantchc}
\cC(m,n)=\{Z=[Z_{ij}]_{i,j\in[m]}\in\rH_{mn,+},Z_{ij}\in\C^{n\times n},\tr Z_{ij} =\delta_{ij}, i,j\in[m]\},\\
\cD(m,n)=\{Z=[Z_{ij}]_{i,j\in[n]}\in\rH_{mn,+},Z_{ij}\in\C^{m\times m}, i,j\in[n], \sum_{i=1}^n Z_{ii}=I_m\}.
\label{unitchc}
\end{eqnarray}
Then $\cC(m,n)$ and $\cD(m,n)$ are compact convex sets of dimension $(mn)^2-m^2=m^2(n^2-1)$.
The map $L\mapsto Z(L)$ is an isomorphism of $\cL_{m,n}$ and $\cC(m,n)$.
In particular, each quantum channel $L$ has a decomposition  \eqref{defcp} such that the conditions \eqref{orthcond} and  \eqref{trprescon}
hold.  Furthermore, if $\tr A_i^*A_i\ne \tr A_j^*A_j$ for $i\ne j$ then the representation \eqref{defcp} of $L$, where the conditions \eqref{orthcond}
hold, is unique up to the change \eqref{trivdefLX} and the order in the summation in \eqref{defcp}.
Similar to the isomorphism $L\mapsto Z(L)$ there is an isomorphism from
$\cL_{m,n}^\vee$ onto $\cD(m,n)$.
\end{theo}
\proof
Let $\cC(m,n)$ be given by \eqref{quantchc}.  Clearly, $\cC(m,n)$ is closed.  Since for any $T\in\cC(m,n)$ the trace of $T$ is $m$ and $T\ge 0$ it follows that
$\cC(m,n)$ is compact.  Let $L:\C^{m\times m}\to \C^{n\times n}$.  By Choi's theorem $Z(L)\in\rH_{mn,+}$ if and only if $L$ is completely positive.
Suppose that $L$ is a completely positive trace preserving transformation.  So $\tr L(\e_i\e_j\trans)=\tr \e_i\e_j\trans=\delta_{ij}$ for $i,j\in [m]$.
Hence $Z(L)\in\cC(m,n)$.  Vice versa, suppose that $Z=[Z_{ij}]_{i,j\in[m]}\in\cC(m,n)$.  Define a linear map $L:\C^{m\times m}\to \C^{n\times n}$
by the equalities $L(\e_i\e_j\trans)=Z_{ij}$ for $i,j\in [m]$.  Clearly, $L:\rH_{m}\to \rH_{n}$.  As $\tr L(\e_i\e_j\trans)=\tr Z_{ij}=\delta_{ij}$ for $i,j\in[m]$
it follows that $L$ is trace preserving.  Choi's theorem yields that $L$ is completely positive.
Hence $L\in\cL_{m,n}$.

Clearly, $\frac{1}{n}I_{mn}\in\cC(m,n)$.  Let
\begin{equation}\label{Wsathomcon}
\cW:=\{W=[W_{ij}]_{i,j\in [m]}\in \rH_{mn}, \quad W_{ij}\in\C^{n\times n},\;\tr W_{ij}=0, \; i,j\in [m]\}.
\end{equation}
Hence $\cW$ is a (real) subspace of $\rH_{mn}$ of dimension $(mn)^2-m^2$.  Note that $Z=\frac{1}{n}I_{nm}+W\in\cC(m,n)$ if and only if
$W\in\cW$ and $\lambda_{mn}(W)$, the smallest eigenvalue of $W$, greater or equal $-\frac{1}{n}$.  Hence $\dim \cC(m,n)=m^2(n^2-1)$.
Other claims for $L$ follow from Theorem \ref{choithm}.

The isomorphism from
$\cL_{m,n}^\vee$ onto $\cD(m,n)$ is proved similarly.
\qed

The next theorem is not new (see Theorem \ref{otcharextpt} and the remarks preceding and following it), but the
proof is new and needed in our characterization of extreme points of $\cL_{m,n}$.
\begin{theo}\label{extptsCmn}  Assume that $L$ is an extreme point of $\cL_{m,n}$.  Then $\rank Z(L)\in [m]$.
\end{theo}
\proof Assume that $L\in\cL_{m,n}$.  Since $L\ne 0$ we must have that $\rank Z(L)\in [mn]$.  Assume that $k=\rank Z(L)>m$.  Let $L$ have the form
\eqref{defcp}, and assume that the conditions
 \eqref{orthcond} hold.  Let $\uu_{k+1},\ldots,\uu_{mn}$ be an orthonormal basis of eigenvectors of $Z(L)$ corresponding to the zero eigenvalue.
Consider the subspace
\begin{equation}\label{subspHmn}
\U(\uu_{k+1},\ldots,\uu_{mn}):=\{U\in\rH_{mn},\;U\uu_i=\0,\quad i=k+1,\ldots,mn  \}.
\end{equation}
We claim the above set is $k^2$ dimensional subspace in $\rH_{mn}$.  Indeed, by assuming that $\uu_1,\ldots,\uu_{mn}$ are standard orthonormal
vectors in $\C^{mn}$ we get that $\U(\uu_{k+1},\ldots,\uu_{mn})=\rH_k\oplus 0$.

The assumption that $k>m$ implies that $\cW\cap \U(\uu_{k+1},\ldots,\uu_{mn})\ne \{0\}$.  Let $U\in \cW\cap \U(\uu_{k+1},\ldots,\uu_{mn})\setminus \{0\}$.
Then for $\varepsilon >0$ small enough $Z(L)\pm \varepsilon U\ge 0$.  Hence $Z(L)\pm \varepsilon U\in \cC(m,n)$.  So there exist $L_1,L_2\in \cL_{m,n}$
such that $Z(L_1)=Z(L)+ \varepsilon U, Z(L_2)=Z(L)- \varepsilon U$.  Since $U\ne 0$ it follows that $L_i\ne L$ for $i=1,2$.  As $\frac{1}{2}(L_1+L_2)=L$
we deduce that $L$ is not an extreme point of $\cL_{m,n}$.  \qed

We now give a necessary and sufficient condition for an extremality of $L\in\cL_{m,n}$.
\begin{theo}\label{necsufconext}  Let $m,n\ge 2$ and assume that $L\in\cL_{m,n}$ and  $k:=\rank Z(L)\le m$.  Then $L$ is an extreme point of $\cL_{m,n}$ if and
only if $\cW\cap  \U(\uu_{k+1},\ldots,\uu_{mn})=\{0\}$, where $ \uu_{k+1},\ldots,\uu_{mn}$ is an orthonormal basis for the null space $Z(L)$.  In particular
for $k=m$ $L$ is an extreme  point if and only if the $m^2$ linear functionals on hermitian matrices of order $mn$ of the form $[Z_{i j}]_{i,j\in[m]}$,
where $Z_{ij}\in\C^{n\times n},i,j\in[m]$,
\begin{equation}\label{m2lincond}
\tr Z_{ii}, \quad \Re \tr Z_{ij}, \quad \Im\tr Z_{ij}, \quad j=i+1,\ldots,m,\quad i\in[m],
\end{equation}
are linearly independent on  $\U(\uu_{m+1},\ldots,\uu_{mn})$.
\end{theo}
\proof  Suppose first that $\cW\cap \U(\uu_{k+1},\ldots,\uu_{mn})\ne\{0\}$.  Then the proof of Theorem \ref{extptsCmn} yields that $L$ is not extreme.
Suppose now that $\cW\cap  \U(\uu_{k+1},\ldots,\uu_{mn})=\{0\}$.  Assume to the contrary that $L$ is not extreme.  So $L=\frac{1}{2}(L_1+L_2), L_1,L_2
\in\cL_{m,n}$ and $L_1,L_2\ne L$.  Then $Z(L)=\frac{1}{2}(Z(L_1)+Z(L_2))$.  Since $Z(L_i)\ge 0$ for $i=1,2$ it follows that $Z(L_i)\in\U(\uu_{k+1},\ldots,\uu_{mn})$.
As $Z(L_i)\in\cC(m,n)$ it follows that $U:=Z(L_1)-Z(L_2)\in\cW\cap \U(\uu_{k+1},\ldots,\uu_{mn})$.  As $U\ne 0$ we obtain a contradiction to our assumption that
$\cW\cap  \U(\uu_{k+1},\ldots,\uu_{mn})=\{0\}$.

Suppose that $k=m$.  Since $\dim\U(\uu_{m+1},\ldots,\uu_{mn})=m^2$ the assumption $\cW\cap  \U(\uu_{m+1},\ldots,\uu_{mn})=\{0\}$ is equivalent
to the assumptions that the $m^2$ linear functionals \eqref{m2lincond} are linearly independent on $\U(\uu_{m+1},\ldots,\uu_{mn})$. \qed

The next result is due to Choi \cite{Cho75}.
\begin{theo} \label{otcharextpt}  Let $L\in\cL_{m,n}$ be of the form \eqref{defcp}.   If the $k^2$ matrices $A_i^*A_j, i,j\in [k]$
are linearly independent then $L$ is an extreme point in $\cL_{m,n}$.  Vice versa, if $L$ is an extreme point of $\cL_{m,n}$ and
$A_1,\ldots,A_k$ are linearly independent, i.e. $\rank Z(L)=k$, then $A_i^*A_j, i,j\in [k]$
are linearly independent.
\end{theo}
\proof  Consider the convex set $\cL_{m,n}^\vee \in \cL(\rH_n,\rH_m)$.   This is the convex set of completely positive unital transformations.
Clearly, $L$ is an extreme point of $\cL_{m,n}$ if and only if $L^\vee$ is an extreme point in   $\cL_{m,n}^\vee$.

Assume that $L$ is of the form \eqref{defcp} such that $A_1,\ldots,A_k$ are linearly independent.
Then $L^\vee$ is given by \eqref{ladj}.
\cite[Theorem 5]{Cho75} implies that $L^\vee$ is extreme if and only if $\{A_i^*A_j\}_{i,j\in[k]}$ are linearly independent.
(See \cite[Remark 4]{Cho75}.)  Observe next that if  $\{A_i^*A_j\}_{i,j\in[k]}$ are linearly independent then $A_1,\ldots,A_k$ are linearly independent.
Combine the above results to deduce the theorem. \qed

Since any  $\{A_i^*A_j\}_{i,j\in[k]}$ for $k>m$  are linearly dependent  in $\C^{m\times m}$, the above theorem yields Theorem \ref{extptsCmn}.
This is essentially pointed out in \cite[Remark 6]{Cho75}, see also \cite{RSW, Rus07}.


%
\section{Extreme points of quantum channels II}
Let
\begin{equation}\label{defLmnk}
\cL_{m,n,k}:=\{L\in\cL_{m,n},\;\rank Z(L)\le k\}.
\end{equation}
The aim of this section is to show that most $L\in\cL_{m.n,k}$, where $k\in \{\lceil\frac{m}{\min(m,n)}\rceil,\ldots,m\}$, are
extreme points of $\cL_{m,n,k}$.
For that we need to recall a few notions of algebraic and semi-algebraic geometry in $\R^N$.
A good reference on semi-algebraic geometry is \cite{Cos05}.

Denote by  $\R[\R^N]$ the ring of polynomials with $N$ variables over $\R$.  For $p\in\R[\R^N]$ denote
\begin{eqnarray*}
Z(p):=\{\x\in\R^N,\; p(\x)=0\}, \quad P_+(p):=\{\x\in \R^N,\;p(\x) >0\}.
\end{eqnarray*}
Assume that $p_1,\ldots,p_k,q_1,\ldots,q_l\in \R[\R^N]$.
An algebraic variety  in $V\subseteq\R^N$ is the set $\cap_{i\in[k]} Z(p_i)$.  A semi-algebraic set is a finite union of basic sets,
where each basic set is $(\cap_{i\in[k]} Z(p_i))\cap (\cap_{j\in l} P_+(q_j))$.
Let $O\subset \R^M$ be a semi-algebraic set.  The image of a semi-algebraic set by a polynomial map $F:\R^N\to R^M$ is semi-algebraic.
If $O,O_1\subset \R^N$ are two semi-algebraic sets than $O\cup O_1,O\setminus O_1$ are semi-algebraic.
We call a semi-algebraic set $O_1\subset O$ a \emph{strong} subset of $O$ if the closure of $O\setminus O_1$ in the standard topology
contains $O$.

 Let $\z_j=\x_j+\ii\y_j,\x_j,\y_j\in\R^M,  j\in[K]$ be an orthonormal set in $\C^M$.
The orthonormality condition is equivalent to the following system of quadratic equations.
\begin{equation}\label{quadcond}
\x_j\trans\x_j+\y_j\trans\y_j=1, \; \x_j\trans\x_l+\y_j\trans\y_l=\x_j\trans\y_l-\y_j\trans\x_l=0, \quad l=j+1,\ldots,K,\;j\in[K].
\end{equation}
Hence, the set of $K$ orthonormal vectors in $\C^M$ is a real algebraic variety in $\R^{2kM}$.

Observe next that $\rH_M$ can be identified with $\R^{M^2}$.  Hence the set $\rH_{M,+}$ is a semi-algebraic set given by the inequalities:
all principal minors of $X\in\rH_{M}$ are nonnegative.  The set of all nonnegative definite matrices of order $M$ of rank $k$ at most, denoted by $\cR(M,k)$
is a semi-algebraic set given by the conditions: all principal minors of order $r$ are nonnegative if $r\le k$ and are zero if $r>k$.
The set $\cR(M,k)^o\subset \cR(M,k)$ of all nonnegative definite hermitian matrices of order $M$ of rank $k$ is an open semi-algebraic set in $\cR(M,k)$.
(It is given by the condition: the sum of all principal minors of order $k$ is positive.)
That is, for each $X\in\cR(M,k)^o$ there exists an open ball $O\in\rH_M$ centered at $X$ such that $O\cap \cR(M,k)\subset\cR(M,k)^o$.
Note that $\cR(M,k)^o$ is an open manifold, which is an open orbit under the action of a general complex linear group of order $M$, which can be identified with a subgroup
of the real general linear group of order $2M$.  It is well known the dimension of $\cR(M,k)^o$ is equal to the dimension of all hermitian matrices of rank $k$ in $\rH_M$,
which is  $k(2M-k)$.
\begin{lemma}\label{nonemptLmnk}   $\cL_{m,n,k}\ne\emptyset$ if and only if
\begin{equation}\label{minkcond}
k\min(m,n)\ge m.
\end{equation}
Furthermore $L\in\cL_{m,n,k}$ if and only if \eqref{defcp} and \eqref{trprescon} hold.
\end{lemma}
\proof  Consider the equality \eqref{trprescon}.  Since  $\rank A_i\le \min(m,n)$ for $i\in [k]$ we deduce \eqref{minkcond}.
Since $\cL_{m,n,k}\subseteq \cL_{m,n,k+1}$ it is enough to show that $\cL_{m,n,k}\ne \emptyset$ for $k=\lceil\frac{m}{\min(m,n)}\rceil$.
Suppose first that $n\ge m$.  So $k=1$ .  Let $U\in\C^{m\times m}$ be a unitary matrix.  Set $A_1^*=[U\;0]$.  Then $A_1^* A_1=I_m$
and $\cL_{m,n,k}$ is nonempty.  Assume now that $1\le n<m$.  Let $\x_1,\ldots,\x_m$ be an orthonormal basis in $\C^m$.
Let $A_i^*=[\x_{(i-1)n+1}\;\ldots\x_{in}]\in\C^{m\times n}$ for $i\in[k-1]$ and $A_k^*=[\x_{(k-1)n+1},\ldots,\x_m,\;0]$.
Then \eqref{trprescon} holds.

Suppose that $L\in\cL_{m,n,k}$, where $k$ satisfies \eqref{minkcond}.   Let $\rank Z(L)=l\le k$.
Then  \eqref{defcp} and \eqref{trprescon} hold, where we can assume that $A_i=0$ for $i>l$.
Vice versa if  \eqref{defcp} and \eqref{trprescon} hold then $\rank Z(L)\le k$.
 \qed

The following lemma is needed to prove the main result of this section.
\begin{lemma}\label{gencond} Let $m,n,k\in\N$ and assume that \eqref{minkcond} holds.  Let $B_1,\ldots,B_k\in \C^{n\times m}$.
Denote
\begin{equation}\label{defBblock}
B=\left[\begin{array}{c}B_1\\ \vdots\\ B_k\end{array}\right]\in \C^{nk\times m}.
\end{equation}
View $B=X+\ii Y$, where $X,Y\in\R^{nk\times m}$ as a real point in $\R^{2nkm}$.
Then
\begin{enumerate}
\item
\begin{equation}\label{poscondB}
B^*B=\sum_{i=1}^k B_i^* B_i >0
\end{equation}
if and only if the matrix $B$ has rank $m$.   In particular all matrices $B$ of the form \eqref{defBblock} which do not satisfy \eqref{poscondB}
form a real algebraic variety in $\R^{2nkm}$.
\item Assume that $k\le m$.  Then the set of $B$'s of the form \eqref{defBblock} such that the $k^2$ matrices $B_i^*B_j, i,j\in [k]$ are linearly
dependent is a real algebraic variety in  $\R^{2nkm}$.
\item  Assume that $k\le m$.   Let $A=[A_1\trans\;\ldots\; A_k\trans]\trans\in \C^{m\times nk}$.  Suppose that $A^* A=I_m$, i.e.
$A_1,\ldots,A_k\in \C^{n\times m}$ satisfy \eqref{trprescon}.  Then there exits a sequence of matrices $A^{(l)}
=[A_{1,l}\trans\;\ldots\;A_{k,l}\trans]\trans\in \C^{m\times nk}$ for $l\in\N$ such that the following conditions hold.
\begin{eqnarray}\notag
&&(A^{(l)})^* A^{(l)}=\sum_{i=1}^k A_{i,l}^*A_{i,l}=I_m,\\
\label{seqAil}
&& A_{i,l}^*A_{j,l} \textrm{ for }i,j\in[k]\textrm{ are linearly independent},\\
&&\lim_{l\to\infty} A_{i,l}=A_i, i\in[k].\notag
\end{eqnarray}
\end{enumerate}
\end{lemma}
\proof \emph{1}.  Clearly, \eqref{poscondB} holds if and only if $B\x=\0\Rightarrow \x=0$.  So \eqref{poscondB} holds if and only if $\rank B=m$.
\eqref{minkcond} yields that $kn\ge m$.  So $\rank B<m$ if and only if all minors of $B$ of order $m$ are zero.  This is an algebraic variety
in the real entries of $X,Y\in\R^{nk\times m}$, where $B=X+\ii Y$.

\emph{2}.   The $k^2$ matrices $B_i^*B_j, i,j\in [k]$ are linearly dependent if an only if the corresponding $k^2$ vectors $\phi_m^{-1}(B_i^*B_j)\in \C^{m^2}$
are linearly dependent.  Let $C=[\phi_m^{-1}(B_1^*B_1)\;\ldots\;\phi_m^{-1} (B_k^* B_k)]\in \C^{m^2\times k^2}$.  Then the $k^2$ matrices
 $B_i^*B_j, i,j\in [k]$ are linearly dependent if and only if all minors of order $k^2$ of $C$ are zero.  To show that this condition is not satisfied for all
$X,Y\in\R^{nkm}$ we need to produce a set of $k$ matrices $B_1,\ldots,B_k\in \C^{n\times m}$ such that the $k^2$ matrices $B_i^*B_j, i,j\in [k]$
are linearly independent.  Let $\y\in\C^n$ be a vector of unit length.  Let $\e_1,\ldots,\e_m\in\R^m$ be the standard orthonormal basis in $\R^m$.
Define $B_i=\y\e_i\trans, i=1,\ldots,k$.  Then $B_i^*B_j=\e_i\e_j\trans$.  Clearly, $B_1^*B_1,\ldots,B_k^* B_k$ are linearly independent.

\emph{3}.  The results of \emph{2} yield that there exist $B_{1,l},\ldots, B_{k,l}$ satisfying the following conditions.
First, $B_{i,l}^* B_{j,l}$ for $i,j\in[k]$ are linearly independent for each $l$.  Second, $\lim_{l\to\infty}B_{i,l}=A_i$ for $i\in [k]$.
Hence we can assume without loss of generality that $D_l:=\sum_{i=1}^k B_{i,l}^*B_{i,l}>0$ for $l\in\N$.
Note that $\lim_{l\to\infty} D_l=I_m$.  Let $F_l\in \rH_{m,+}$ the unique positive square root of $D_l$, i.e. $F_l^2=D_l$.
Clearly, $\lim_{l\to\infty} F_l=I_m$.  Set $A_{i,l}:=B_{i,l}F_l^{-1}$ for $i\in[k]$ and $l\in\N$.  Hence the first condition of \eqref{seqAil} holds.
Similarly,  the third condition of \eqref{seqAil} holds.   As $A_{i,l}^*A_{j,l}=F_l^{-1}(B_{i,l}^* B_{j,l})F_l^{-1}$ for $i,j\in[k]$ we deduce
 the second condition of \eqref{seqAil}.\qed

\begin{theo}\label{genextptm}  Let $m,n,k\in \N$ and assume that $k\in \{\lceil\frac{m}{\min(m,n)}\rceil,\ldots,m\}$.  Then
\begin{enumerate}
\item $\cL_{m,n,k}$ is a semi-algebraic set.  Let $\cL_{m,n,k}'$ be the set of all nonextreme points of $\cL_{m,n}$ which are in
 $\cL_{m,n,k}$.
Then $\cL_{m,n,k}'$ is a strong semi-algebraic subset of $\cL_{m,n,k}$.  In particular,  $\cL_{m,n,k}\setminus\cL_{m,n,k}'$,
the set of extreme points of $\cL_{m,n}$ in $\cL_{m,n,k}$, is a semi-algebraic set which is dense in $\cL_{m,n,k}$.
\item
The set of extreme points $Z\in\cC(m,n)$ of rank $m$ is a nonempty open semi-algebraic set in
$\cC(m,n)\cap\cR(mn,m)^o$, each of whose connected components is of dimension $2m^2(n-1)$.
\end{enumerate}
\end{theo}
\proof \emph{1}.
Let
\[\cC(m,n,k):=\{Z\in \cC(m,n),\;\rank Z\le k\}, \cC(m,n,k)^o:=\{Z\in \cC(m,n),\;\rank Z= k\},\]
for $k\in[mn]$.
Consider all $B\in \C^{nk\times m}$ of the form \eqref{defBblock}.  Then $\sum_{i=1}^k B_i^* B_i=I_m$ if and only if
$B^*B=I_m$.  So the set of all $B\in \C^{nk\times m}$ satisfying $B^*B=I_m$ is an algebraic variety  $V\subset \R^{2nkm}$ given by
quadratic polynomials.
We now define $F:\C^{nk\times m}\to \rH_{mn}$.  Namely
\[F(B)=\sum_{i=1}^k \hat B_i (\hat B_i)^*.\]
Again, this map is a polynomial map, where each coordinate is quadratic.  Clearly, $F(V)=\cC(m,n,k)$.  Hence $\cC(m,n,k)$ is
semi-algebraic.  Let $\cC(m,n,k)'$ be all non-extreme points of $\cC(m,n)$ which are in $\cC(m,n,k)$.
Let $V'\subset V$ be the following strict subvariety of $V$.  Namely, it consists of all $B\in V$, such that the $k^2$ blocks
$B_i^*B_j, i,j\in [k]$ are linearly dependent.  (See \emph{2} of the proof of Lemma \ref{gencond}.)  \cite[Theorem 5]{Cho75} yields that
$F(V')=\cC(m,n,k)'$.  The claim that $\cC(m,n,k)\setminus \cC(m,n,k)'$ is dense in $\cC(m,n,k)$ follows from part \emph{3}
of Lemma \ref{gencond} and   \cite[Theorem 5]{Cho75}.\\

\emph{2}.  Since $\cL_{m,n}$ and $\cC(m,n)$ are isomorphic, (Theorem \ref{charqc}), Lemma
\ref{necsufconext} implies that $X\in\cC(m,n,m)^o$ is an extreme point if and only if
the following condition hold.  Let $\cY(X)\subset\rH_{mn}$ be the set of hermitian matrices whose kernel contains
the kernel of $X$.  Then the $m^2$ linear functionals given by \eqref{m2lincond} on $\cY(X)$ are linearly independent.
We first show that $\cC(m,n,m)^o\ne \emptyset$.  Let $X=[X_{ij}]_{i,j\in[m]}$ be of the form
$X_{ij}=\delta_{ij}(\e_1\e_1\trans), i,j\in [m]$.  Clearly $X\in \cC(m,n,m)^o$. It is straightforward to show that
$$\cY(X)=\{Y=[Y_{ij}]_{i,j\in [m]}\in \rH_{mn}, \;Y_{ij}=a_{ij}(\e_1\e_1\trans), i,j\in [m], \; A=[a_{ij}]\in\rH_m\}.$$
Clearly, the $m^2$ linear functionals \eqref{m2lincond} are linearly independent on $\cY(X)$.

Recall that $\cR(mn,m)^o$ is a manifold of dimension $m^2(2n-1)$. Assume that $X \in \cC(m,n,m)^o$ is an extreme point.
Let $\cT(X)$ be the tangent hyperplane to $\cR(mn,m)^o$ at $X$.  Then $\cY(X)$ is an $m^2$ dimensional subspace of $\cT(X)$.
As the $m^2$ linear functionals given by \eqref{m2lincond} on $\cY(X)$ are
linearly independent it follows that these $m^2$ conditions are linearly independent on $\cT(X)$.
Observe next that $\cC(m,n,m)^o$ locally at  $X$ is a submanifold of $\cR(mn,m)^o$ given by the linear conditions stated in \eqref{quantchc}.
Hence the tangent hyperplane of $\cC(m,n,m)^o$ at $X$
is the subspace of $\cT(X)$ consisting of all $Y=[Y_{ij}]_{i,j\in[m]}\in\rH_{mn}$ such that $\tr Y_{ij}=0$ for $i,j\in [m]$.
Since $X$ is an extreme point it follows that the dimension of the tangent hyperplane of $\cC(m,n,m)^o$ at $X$ is $m^2(2n-1)-m^2$.
As a nonempty open semi-algebraic set in  $\cC(m,n,m)^o$ is a union of a finite number of connected components,
 it follows that each connected component is an open manifold of dimension $2m^2(n-1)$.
\qed

Note that the claim that the closure of all extreme points in $\cL_{m,n}$ is $\cL_{m,n,m}$, i.e. the case $k=m$ of part \emph{1} of
the Theorem \ref{genextptm}, was observed in \cite[Theorem 1]{Rus07}.

The following theorem characterizes the sets of extreme points of $\cL_{m,n}$ which belong to special $\cL_{m,n,k}$.
\begin{theo}\label{charextreptLmnk}  Let $2\le m,n\in\N$.  Denote $k_0:=\lceil\frac{m}{\min(m,n)}\rceil$.
Assume that $k\in\{k_0,\ldots,m\}$.  Then
\begin{enumerate}
\item Suppose that $L\in \cL_{m,n}$ with $\rank Z(L)=k$  is not an extreme point.  Then $L$ is a convex combination of some $L_1,L_2\in \cL_{m,n,k-1}$.
\item Each $L\in\cL_{m,n,k_0}$ is an extreme point of $\cL_{m,n}$.
\item $L\in \cL_{m,n}$ with $\rank Z(L)=k_0+1$ is not an extreme point if and only if it is a convex combination of two distinct extreme points given in 2.
\end{enumerate}

\end{theo}
\proof Assume that $L\in\cL_{m,n}$ and $\rank Z(L)=k$.  Suppose that $L$ is not an extreme point of $\cL_{m,n}$.
So there exist $M_1,M_2\in\cL_{m,n}, M_1\ne M_2$ such that $L=\frac{1}{2}(M_1+M_2)$.  As $Z(M_1),Z(M_2) \ge 0$ it follows that $\rank Z(M_1)\le k$.
Since $2Z(L)\ge Z(M_1)$ it follows that the null space of $Z(L)$ is  contained in the null space of $Z(M_1)$.
Let $W=Z(L)-Z(M_1)$.  So $W=[W_{ij}]_{i,j\in [m]}\in\rH_{mn}$ is nonzero and satisfies the conditions $\tr W_{ij}=0$ for $i,j\in [m]$.
Furthermore, the null space of $W$ contains the null space of $Z(L)$.  Consider the matrix $Z(L)+tW$.  It is straightforward to see that
there exists $\epsilon >0$ such that for each $t\in[-\epsilon,\epsilon]$ the matrix $Z(t):=Z(L)+tW$ is nonnegative definite.  Since $\tr W=0$
it follows that there exists $t_1,t_2>0$ such that $Z(t_1),Z(-t_2)\ge 0$ and $\rank Z(t_1),\rank Z(-t_2) <k $.  Clearly, $Z(t_1),Z(-t_2)\in \cC(m,n)$.
Hence $Z(t_1)=Z(L_1),Z(-t_2)=Z(L_2)$ for corresponding $L_1,L_2\in\cL_{m,n}$.   Furthermore, $L_1,L_2\in \cL_{m,n,k-1}$.
As $L=\frac{t_2}{t_1+t_2}M_1+\frac{t_1}{t_1+t_2}M_2$ we deduce \emph{1}.

Lemma \ref{nonemptLmnk} yields that $\cL_{m,n,k_0-1}=\emptyset$.  Assume that $L\in\cL_{m,n,k_0}$.  Hence $\rank Z(L)=k_0$.
Part \emph{1} yields that $L$ is an extreme points.  This proves \emph{2}.

Suppose that $L\in\cL_{m,n,k_0+1}$ and $\rank Z(L)=k_0+1$.  If $L$ is a convex combination of $L_1,L_2\in \cL_{m,n,k_0}, L_1\ne L_2$ then $L$ is not an
extreme point of $\cL_{m,n}$.  Assume now that $L$ is not an extreme point.  Combine \emph{1-2} to deduce that $L$ is a convex combination
of two extreme points given in \emph{2}.\qed

\begin{defn}\label{defredqc}  Let $2\le m,n\in\N$.  $L\in\cL_{m,n}$ is called decomposable if the following conditions are satisfied.
There exist two unitary matrices $U\in \C^{m\times m}, V\in \C^{n\times n}$
and $p,q\in\N, p+q=m$ with the following properties.  Let $\tilde L\in\cL_{m,n}$ be given as $\tilde L(X)=VL(UXU^*)V^*$.  Then
there exists $L_1\in \cL_{p,n},L_2\in\cL_{q,n}$ such that
\begin{eqnarray}\label{defredqc1}
&&\tilde L(\left[\begin{array}{cc} X_{11}&X_{12}\\X_{21}&X_{22}\end{array}\right])=L_1(X_{11})+L_2(X_{22}),\\
&&X_{11}\in \C^{p\times p},X_{12}\in\C^{p\times q}, X_{21}\in \C^{q\times p}, X_{22}\in\C^{q\times q}.  \notag
\end{eqnarray}
Equivalently, $Z(\tilde L)$ is a block diagonal matrix $\diag(Z(L_1),Z(L_2))$.
We denote $\tilde L$ by $L_1\oplus L_2$ if the above equality holds.
\end{defn}
(It is straightforward to see that $\tilde L$ is indeed quantum channel if $L_1,L_2$ are quantum channels.)

The following Lemma gives sufficient conditions for a decomposable quantum channel to be a nonextreme point in $\cL_{m,n}$.
\begin{lemma}\label{nextredc}  Let $L\in\cL_{m,n}$ be a decomposable channel as defined in Definition \ref{defredqc}.
Then $L$ is not an extreme point if one of the following conditions hold.
\begin{enumerate}
\item $L_1$ is a nonextreme point in $\cL_{p,n}$.
\item $L_2$ is a nonextreme point in $\cL_{q,n}$
\item $\rank Z(L_1)\rank Z(L_2)>pq$.
\end{enumerate}
\end{lemma}
\proof  Suppose that $L_1$ is nonextreme.  So $L_1=\frac{1}{2} (M+N)$, where $M, N\in\cL_{p,n}, M\ne N$.  Then $\tilde L=\frac{1}{2}(M\oplus L_2+N\oplus L_2)$.
Hence $\tilde L$ and $L$ are not extreme.  This proves \emph{1}.  \emph{2} is proved similarly.

We now show \emph{3}.
Assume that $\rank Z(L_1)\rank Z(L_2)>pq$.  So either $\rank Z(L_1)>p$ or $\rank Z(L_2)>q$.   Theorem \ref{extptsCmn} yields that
either $L_1$ or $L_2$ are nonextreme.  Part \emph{1} or \emph{2} yield that $L$ is nonextreme.  \qed

\section{A condition on the image of $L\in\cL_{m,n,k}$}
\begin{theo}\label{condimLmnk}
Let $m,n,k\in \N$.  Assume that $m,n\ge 2,k\ge 1$ be integers.  Suppose that  $L$ is of the form \eqref{defcp}, where $A_i\ne 0$ for $i\in[k]$.
Let $l=\dim\span(A_1,\ldots,A_k)$.  Assume for simplicity of notation that $A_1,\ldots,A_l$ are linearly independent.


Define
\begin{equation}\label{defBrA}
B_j:=[A_1\e_j\;\ldots A_l\e_j]\in \C^{n\times l},\; j\in[m],\quad r=\dim\span(B_1,\ldots,B_m).
\end{equation}

Then $L(\cP_m)$ contains a nonzero matrix of rank  at most $\min(n,l)$.  Moreover, if $l\ge 2$ then

\begin{enumerate}
\item Each matrix in $L(\cP_m)$ has rank at most $\min(n,l)$.

\item $L\in\cL_{m,n}$ implies that $r=m$.

\item If  $r\ge \max(n,l)-\min(n,l)+2$ then $L(\cP_m)$ contains a nonzero matrix of rank strictly less than $\min(n,l)$.
More precisely,  let $p$ be the smallest positive integer satisfying
\begin{equation}\label{pcond}
p(n+l-p)+r\ge nl+1.
\end{equation}
Then $L(\cP_m)$ contains a nonzero matrix of rank $p$ at most.
In particular, $L(\cP_m)$ contains a matrix of rank one if $r\ge (n-1)(l-1)+1$.
\end{enumerate}
\end{theo}
\proof  Clearly, the range of $Z(L)$ is spanned by $\hat A_1,\ldots,\hat A_k$.  Hence $\rank Z(L)=l$.
So $L(X)=\sum_{i=1}^l C_i X C_i^*$ for some linearly independent $C_1,\ldots,C_l\in \C^{n\times m}$.  Therefore

\noindent
$L(\uu\uu^*)=\sum_{i=1}^l (C_i\uu)(C_i\uu)^*$.  So $\rank L(\uu\uu^*)\le l$.  Since $L(\uu\uu^*)\in \C^{n\times n}$
it follows that $\rank L(\uu\uu^*)\le n$.  This establishes \emph{1}.

For $\x=(x_1,\ldots,x_m)\trans\in\C^m$ let $M(\x)=[A_1\x\ldots A_l\x]\in \C^{n\times l}$.
Observe first that $\y^*L(\x\x^*)=0\iff \y^*M(\x)=\0$.  Hence $\rank L(\x\x^*)=\rank M(\x)$.
Observe next that $M(\x)=\sum_{i=1}^m x_i B_i$.

Assume that $L\in\cL_{m,n}$.  Then $\tr L(\x\x^*)=\x^*\x$.
Suppose that $B_1,\ldots,B_m$ are linearly dependent.  So the there is $\x\ne \0$ so that $0=\sum_{i=1}^m x_i B_i=M(\x)$.
Hence $L(\x\x^*)=0$ which is impossible.  This establishes \emph{2}.

Let $\Phi(n,l,p)\subset \C^{n\times l}$ be the complex variety of matrices of rank at most $p\;(<\min(n,l))$.  It is well known that this is an irreducible variety
of complex dimension $p(n+l-p)$.  (Choose first $p$ rows linearly independent, and all other rows to be linear combinations of the first $p$ rows.)
Let $\Psi\subset \C^{n\times l}$ be a subspace of complex dimension $d$.  Then $\Phi(n,l,p)\cap \Psi\ne \{0\}$ if $p(n+l-p)+d\ge nl+1$.
Note that the set of all $M(\x)$ for $\x\in\C^m$ is a subspace of dimension $r$.  Clearly $(\min(n,l)-1)( n+l-\min(n,l)+1)+r\ge nl+1$
if and only if $r\ge \max(n,l)-\min(n,l)+2$.  Furthermore, there  exists a nonzero $M(\x)$ of rank at most  $p$, where $p$ is the minimal solution of \eqref{pcond}.
Finally $1\cdot(n+l-1)+r\ge nl+1$ if and only if $r\ge (n-1)(l-1)+1$.   This establishes \emph{3}.
 \qed

\begin{corol}\label{existpurest22}  Each $L\in\cL_{m,2,m}$ maps some pure state to a pure state.
\end{corol}

For $m=2$ this result follows from \cite[Theorem 16]{RSW}.
\section{Extreme points of $\cL_{2,2}$}
The results of this section essentially appear also in \cite{RSW}, but the approach there is different.
\begin{theo}\label{rankhL2}  Let  $L$ be a quantum channel in $\cL_{2,2}$ such that $\rank Z(L)=2$.  Then one of the following conditions hold.
\begin{enumerate}
\item\label{im1pt} $L(\rH_{2,+,1})=\{R\}$, where $R$ is rank one hermitian matrix of trace one.
\item\label{imint} $L(\rH_{2,+,1})$ is the convex hull of two distinct rank one hermitian matrices $R_1,R_2$, called an interval $[R_1,R_2]$.
More precisely,  there exist two orthogonal rank one matrices $D_1,D_2\in\rH_{2,+,1}$ such that $L(D_i)=R_i, i=1,2$ and $L(T)=0$ for any
$T\in\rH_2$ which is orthogonal to $D_1,D_2$.
\item\label{imrnk1mat}  $L(\rH_{2,+,1})$ contains exactly two distinct rank one matrices $R_1,R_2$ and strictly contains the interval $[R_1,R_2]$.
\item $L(\rH_{2,+,1})$ contains exactly one rank one matrix $R$, whose preimage is a unique rank one matrix $D\in\rH_{2,+,1}$.
\end{enumerate}
\end{theo}
\proof   Corollary \ref{existpurest22} implies that there exists a rank one matrix $A\in\rH_{2,+,1}$ so that $L(A)$ is a rank one matrix.
In what follows we assume that $L$ has the property $L(\e_1\e_1\trans)=\e_1\e_1\trans$.  (This can always be achieved by replacing $L$ with $L_1$, where
$L_1(X)=V^*L(U^*XU)V$ for some unitary $U,V\in C^{2\times 2}$.)
The assumption that $Z(L)\in\rH_{4,+}$, $\tr L(\e_1\e_2\trans)=0$ and $\rank Z(L)=2$ yields that
\begin{equation}\label{formhatL1}
Z(L)=\left[\begin{array}{cccc}1&0& 0&y\\0&0&0&0\\0&0 &1-c& s\\
\bar y&0& \bar s& c
\end{array}\right],\; |y|^2\le c\le 1, (1-c)(c-|y|^2)=|s|^2.
\end{equation}

Suppose first that $y=0$, i.e. $L(\e_1\e_2\trans)=0$.  Then $L(\e_2\e_2\trans)$ is a rank one matrix.  If $L(\e_2\e_2\trans)=\e_1\e_1\trans$
we are in the case \emph{1} of our theorem.  If $L(\e_2\e_2\trans)\ne\e_1\e_1\trans$
we are in the case \emph{2} of our theorem.

Assume now that $y\ne 0$.  Since we assume that $\rank Z(L)=2$ it follows that $|y|<1$.
We now discuss the conditions when $L(\rH_{2,+,1})$ contains a rank one matrix $C$ different from $\e_1\e_1\trans$.
Assume that $L(B)=C, B\in\rH_{2,+,1}$.  So $B$ is a solution to the minimal problem discussed in beginning of the proof.
If $\rank B=2$ then $B$ is a convex combination of two rank one matrices $B_1,B_2\in\rH_{2,+,1}$.  Since $\lambda_2(\cdot)$ is a concave
function on $\rH_2$ it follows that $\rank L(B_1)=\rank L(B_2)=1$.  Hence we deduce that if $L(\rH_{2,+,1})$ contains at least two different rank one matrices
$L(B)$ is a rank one matrix for some $B$ of the form $B=(v,\bar w)\trans (\bar v,w), |v|^2+|w|^2=1,w\ne 0$.
Observe
\[L(B)=\left[\begin{array}{cc} |v|^2+|w|^2(1-c)& v w y+|w|^2 s\\ \bar v \bar w\bar y +|w|^2\bar s& |w|^2c\end{array}\right].\]

Suppose first that $c=1$.  Then $s=0$ and $\det L(B)>0$ for $v\ne 0$.  For $v=0$ we obtain that $B=\e_2\e_2\trans$.
Then $L(\e_2\e_2\trans)=\e_2\e_2\trans$.
This corresponds to the case \emph{3} of our theorem.  (Note that here $R_1$ and $R_2$ are orthogonal.)

Assume now that $c<1$.  Then $\det L(\e_2\e_2\trans)=(1-c)c-|s|^2=(1-c)|y|^2>0$, so $\rank  L(\e_2\e_2\trans)=2$.
So it is enough to consider the above rank one matrix $B$ with $v,w\ne 0$.
Taking into account the equality  $(1-c)(c-|y^2|)=|s|^2$ we deduce.
\begin{eqnarray*}
&&\det L(B)=|v w|^2(c-|y|^2 + |w'|^2(1-c)c -|w'|^2|s|^2-2\Re(w'y\bar s))=\\
&&|v w|^2(\frac{|s|^2}{1-c}+(1-c)|u|^2-2\Re(u\bar s))=\\
&&|v w|^2 (1-c)|u-\frac{s}{1-c}|^2, \quad  w'=\frac{w}{\bar v},u=w' y.
\end{eqnarray*}
If $s=0$, i.e. $c=|y|^2$, then $\det L(B)>0$.  This corresponds to the case \emph{4} of our theorem.

Suppose that $s\ne 0$.  Then  $L(B)$ has rank one for a unique value of $w'=\frac{s}{y(1-c)}$.
Since $|w|<1$ we deduce that $|w|^2 c<c$.  Hence $L(\e_2\e_2\trans)$ can not be a convex combination of $R_1=\e_1\e_1\trans$ and $R_2=L(B)$.
This corresponds to case \emph{3} of our
theorem.  \qed

Theorem \ref{charextreptLmnk} yields.
\begin{corol}\label{extptrhl2}  Let $L\in\cL_{2,2}$.  Then $L$ is an extreme point of $\cL_{2,2}$ if and only if one of the following conditions hold.
\begin{enumerate}
\item $L(X)=UXU^*$ for some unitary matrix $U$, i.e. $\rank Z(L)=1$.
\item $\rank Z(L)=2$ and $L$ is not a convex combination of two extreme points given in \emph{1}.
\end{enumerate}
\end{corol}

\begin{corol}\label{sufcondext}  Let $L\in\cL_{2,2,2}$ and and assume that $L$ is not unital.  Then $L$ is an extreme point of $\cL_{2,2}$.
\end{corol}

\section{Extreme points of $\cL_{3,2}$}
In this section we characterize the extreme points of $\cL_{3,2}$.
\begin{defn}\label{defnMt}  For $t\in (0,1]$ denote by $\cM_t$ all completely positive $M:\C^{2\times 2}\to \C^{2\times 2}$ satisfying
\begin{equation}\label{extptLt1}
\tr M(\f_1\f_1\trans)=1,\; \tr M(\f_2\f_2\trans)=t,\; \tr M(\f_1\f_2\trans)=0, \; \f_1=(1,0)\trans,\;\f_2=(0,1)\trans.
\end{equation}
\end{defn}
Note that $\cM_1=\cL_{2,2}$.
For completeness we characterize the extreme points of $\cM_t$, although this characterization will not be used in the proof of
Theorem \ref{extrmpts32}.
The proof of the following lemma follows straightforward from the arguments of the proof of Theorem \ref{charextreptLmnk}.
\begin{lemma}\label{extptLt} For  $t\in (0,1]$ the set $\cM_t$ is compact and convex.  $M$ is an extreme points if and only if one of the following conditions hold.
\begin{enumerate}
\item $\rank Z(M)=1$, i.e., $M(X)=AXA^*$, where $A\in \C^{2\times 2}$ and $A^*A=\diag(1,t)$.
\item $\rank Z(M)=2$ and $M$ is not a convex combination of two extreme points given by 1.
\end{enumerate}
\end{lemma}
The following lemma is well known and we bring its proof for completeness.
\begin{lemma}\label{rangelem}  Let $A\in\rH_{m,+}, \uu\in\C^{m}\setminus\{\0\}$.  Then
\begin{enumerate}
\item The range of $B:=A+\uu\uu^*$ contains $\uu$.
\item Assume that the range of $A$ contains $\uu$.  Then there exists $\epsilon >0$ such that $A-\epsilon \uu\uu^*\ge 0$.
\item Assume that $A$ has the following block form $A=\left[\begin{array}{cc}1&\vv^*\\ \vv&D\end{array}\right]$, where
$\vv\in\C^{m-1}, D\in\rH_{m-1,+}$.  Then
\begin{equation}\label{nonnegADv}
 \quad D-\vv\vv^*\ge 0.
\end{equation}
\end{enumerate}
\end{lemma}
\proof
\emph{1}.
If the range of $A$ contains $\uu$  then the range of $B$ contains $\uu$.  Suppose that the range of $A$ does not contain $\uu$.
Let $P$ be the orthogonal projection on the range of $A$.
Clearly $B(I-P)\uu=\uu(\uu^*(I-P)\uu)\ne \0$.  So $\uu$ in the range of $B$.

\emph{2}.  By restricting $A$ to its range it is enough to assume that $A>0$.  Then  $A-\epsilon \uu\uu^*\ge 0$ for some $\epsilon>0$.

\emph{3}.  Clearly,
\[\left[\begin{array}{cc}1&0\\ 0&D-\vv\vv^*\end{array}\right]=
\left[\begin{array}{cc}1&0\\ -\vv&I_{m-1}\end{array}\right]
\left[\begin{array}{cc}1&\vv^*\\ \vv&D\end{array}\right]
\left[\begin{array}{cc}1&-\vv^*\\ 0&I_{m-1}\end{array}\right].\]
Hence \eqref{nonnegADv} holds.
\qed

\begin{theo}\label{extrmpts32}  Let $L\in \cL_{3,2,3}$.  Then there exist pure states $B\in\rH_{3,+,1},C\in \rH_{2,+,1}$  such $L(B)=C$.
By choosing orthonormal bases $\e_1,\e_2,\e_3\in\C^3, \f_1,\f_2\in\C^2$  such that $B=\e_1\e_1^*, C=\f_1\f_1^*$ we can assume that
\begin{equation}\label{sepffrmL32}
Z(L)=\left[\begin{array}{cccccc}1&0&0&x&0&y\\0&0&0&0&0&0\\0&0&1-c&s&a&b\\  \bar x&0&\bar s&c&d&-a\\0&0&\bar a&\bar d&1-e&f\\
\bar y&0&\bar b&-\bar a&\bar f&e\end{array}\right]=[W_{ij}]_{i,j\in[3]},\; W_{ij}\in\C^{2\times 2},i,j\in[3],
\end{equation}
where $c,e\in[0,1]$ and $|x|^2\le c, |y|^2\le e$.
Furthermore, we can assume that $x=0$.
Let $L_2\in \cL_{2,2}$ be defined by
\begin{equation}
Z(L_2):=\left[\begin{array}{cccc}1-c&s&a&b\\\bar s&c&d&-a\\ \bar a&\bar d&1-e&f\\\bar b&-\bar a&\bar f&e\end{array}\right].
\end{equation}
Let $\g_i:=(\delta_{i1},\ldots,\delta_{i4})\trans, i\in[4]$ be the standard basis in $\C^4$.
\begin{enumerate}
\item $|y|=1$.  Then $e=1,a=b=d=f=0$.  $L$ is extreme if and only if $W_{22}$ has rank one.
\item $|y|\in [0,1)$.   Let $M\in\cM_{1-|y|^2}$ be given by $Z(M)=Z(L_2)-|y|^2\g_4\g_4\trans$.  Then
$\rank Z(M)\le 2$.  $L$  is extreme if and only if  either  $\rank Z(M)=1$ or $\rank Z(M)=2$, $M$ is an extreme point in $\cM_{1-|y|^2}$ and one of the following conditions hold.
\begin{enumerate}
\item $y=0$ and the range of $Z(L_2)$ does not contain a nonzero vector of the form $z\g_2+w\g_4$.
\item $|y|\in (0,1)$ and there does not exist $(z,w)\trans\ne \0$ such that the following two inequalities hold.
\begin{eqnarray}\notag
Z(L_2)\ge (z\g_2+(y+w)\g_4) (z\g_2+(y+w)\g_4)^*,\\
\label{twoforbidin}\\
\quad Z(L_2)\ge (-z\g_2+(y-w)\g_4) (z\g_2+(y-w)\g_4)^*.\notag
\end{eqnarray}
\end{enumerate}
\end{enumerate}

\end{theo}
\proof   Assume that $L\in\cL_{3,2,3}$.
Observe first that there is no $A\in \C^{2\times 3}$ such that $A^* A=I_3$.  Hence $\rank Z(L)\in\{2,3\}$.
So the value of $l$ in Theorem \ref{condimLmnk} is either $2$ or $3$.   Corollary \ref{existpurest22}
yields that there are pure states
$B\in\rH_{3,+,1}, C\in\rH_{2,+,1}$ so that $L(B)=C$.  Choose orthonormal bases $\e_1,\e_2,\e_3$ and $\f_1,\f_2$ in $\C^3$ and $\C^2$ respectively
so that $A=\e_1\e_1^*, B=\f_1\f_1^*$.  Assuming that $\e_1,\e_2,\e_3$ and $\f_1,\f_2$ are the standard bases in $\C^3$ and $\C^2$ respectively
we obtain that $Z(L)$ is of the form \eqref{sepffrmL32}.   Hence $c,e\in [0,1]$.
Since $Z(L)\ge 0$ part \emph{3} of Lemma  \ref{rangelem}
yields that $Z(L)$ is congruent to $\f_1\f_1\trans\oplus N$, where
\begin{equation}\label{defmatN}
N:=Z(L_2)-(\bar x\g_2+\bar y\g_4)(\bar x\g_2+\bar y\g_4)^*\ge 0.
\end{equation}
Since the diagonal entries of $N$ are nonnegative we deduce that $|x|^2\le c, |y|^2\le e$.

We now show that by changing an orthonormal basis in $\C^3$ to $\e_1,\tilde\e_2, \tilde\e_3$
we may assume that $x=0$.  Suppose that $x\ne 0$.  Let $\w:=-\bar y\e_2+\bar x\e_3$. Then
\[L(\e_1\w^*)=L(-y\e_1\e_2^*+x\e_1\e_3^*)=-y\left[\begin{array}{cc}0&x\\0&0\end{array}\right]+x\left[\begin{array}{cc}0&y\\0&0\end{array}\right]=0.
\]
Let $\tilde \e_2:=\frac{1}{\|\w\|}\w$ and $\tilde\e_3$ be a unit vector orthogonal to $\e_1,\tilde\e_2$.  Assuming that $\e_1,\tilde\e_2,\tilde\e_3$
is now the standard basis in $\C^3$ we deduce that  $Z(L)$ has the form \eqref{sepffrmL32} with $x=0$.

\emph{1}.  Clearly,  $|y|\le \sqrt{e} \le 1$.
Assume that $|y|=1$.  Then $e=1$ and $1-e=0$.   As $N\ge 0$ it follows that
$e=1,a=b=d=f=0$.
Let $\h_i=(\delta_{i1},\ldots,\delta_{i6})\trans,i\in [6]$ be the standard basis in $\C^6$.
So
$$Z(L)=(\h_1+\bar y\h_6)(\h_1+\bar y\h_6)^* +\diag(0, W_{22}, 0), \quad  0,W_{22}\in \rH_{2,+,1}.$$
If $W_{22}$ is a rank one matrix then $\rank Z(L)=1+\rank \diag(0, W_{22}, 0)=2$ and $L$ is extreme.
Suppose $\rank W_{22}=2$.  So $W_{22}=t U_1+(1-t)U_2, t\in(0,1)$ for some distinct rank one matrices $U_1,U_2\in\rH_{2,+,1}$.
Then $L$ is nonextreme since $Z(L)$ equals to:
$$t((\h_1+\bar y\h_6)(\h_1+\bar y \h_6)^* +\diag(0, U_1, 0))+(1-t)((\h_1+\bar y\h_6)(\h_1+\bar y\h_6)^* +\diag(0, U_2, 0)).$$
This proves part \emph{1}.

\emph{2}.  Assume now that $|y|\in[0,1)$.  Let $N$ be defined by \eqref{defmatN}, where $x=0$.
Then $N=Z(M)$ for some $M\in\cM_{1-|y|^2}$.
If $\rank N=1$ then $\rank Z(L)=2$ and $L$ is extreme.  Suppose now that $\rank N=2$.
Assume $M$ is nonextreme in $\cM_{1-|y|^2}$.  Let $M=\frac{1}{2}(M_1+M_2), M_1,M_2\in\cM_{1-|y|^2}, M_1\ne M_2$.
So  $Z(M)=\frac{1}{2}(Z(M_1)+Z(M_2))$.  Let $P_i:=Z(M_i)+|y|^2\g_4\g_4\trans, i=1,2$.  Clearly $P_1,P_2\in\cC(2,2)$.
Denote $Q_i\in \rH_{6,+}$  the matrix obtained from $Z(L)$ be replacing the submatrix  $Z(L_2)$ by $P_i$ respectively.
Lemma  \ref{rangelem} yields that $Q_1,Q_2\in\rH_{6,+}$.  Clearly, $Q_1,Q_2\in \cC(3,2)$ and $Q_1\ne Q_2$.  Hence there exists $\tilde L_1,
\tilde L_2\in\cL_{3,2}$
such that $Z(\tilde L_i)=Q_i$ for $i=1,2$.  By construction $Z(L)=\frac{1}{2}(Q_1+Q_2) $.  Hence $L=\frac{1}{2}(\tilde L_1+\tilde L_2)$
and $L$ is nonextreme.

Assume now that $M$ is an extreme point in $\cM_{1-|y|^2}$.

\emph{a}.  Suppose that $y=0$.  Then $N=Z(L_2)$.
Assume first that the range of $N$ contains a nonzero vector $z\g_2+w\g_4$.
Lemma  \ref{rangelem} yields that there exists $\epsilon\in(0,1)$ such that $N_1:=N-\epsilon^2(z\g_2+w\g_4)(z\g_2+w\g_4)^*\ge 0$.
Let $G_1,G_2$ be obtained from $Z(L)$ by replacing the entries $(1,4),(1,6),(4,1),(6,1)$  by the following entries respectively
\begin{equation}\label{defG1G2}
\begin{array}{ccccc}
 &\epsilon \bar z,\quad & y+\epsilon \bar w,\quad &\epsilon z, \quad &\bar y+\epsilon w\\
 &-\epsilon \bar z,\quad &y-\epsilon \bar w,\quad &-\epsilon z, \quad  &\bar y-\epsilon  w.
\end{array}
\end{equation}
(Recall that $y=0$.)  Since $N_1\ge 0$ Lemma \ref{rangelem} that $G_1,G_2\in\cC(3,2)$.
As $Z(L)=\frac{1}{2}(G_1+G_2)$ we deduce
that $Z(L)$ is nonextreme in $\cC(3,2)$.   Hence $L$ is nonextreme.

Vice versa, assume that $L$ is nonextreme.  So $L=\frac{1}{2}(\tilde L_1+\tilde L_2)$ for some $\tilde L_1,\tilde L_2\in\cL_{3,2}$.
Let $G_i=Z(\tilde L_i)$ for $i=1,2$.  Since $Z(L_2)$ was extreme in $\cM_1$ it follows that $G_1,G_2$ are obtained from $Z(L)$
by replacing the entries $(1,4),(1,6),(4,1),(6,1)$ with the entries given by \eqref{defG1G2}, where $y=0,\epsilon=1$ and $(z,w)\trans\ne \0$.
Hence $Z(L_2)-(z\g_2+w\g_4)(z\g_2+w\g_4)^*\ge 0$.  Therefore the range of $Z(L_2)$ contains a nonzero vector $z\g_2+w\g_4$.

\emph{b}.  Assume now that $|y|\in (0,1)$.  Suppose first that there exists $(z,w)\trans\ne \0$ such that the two inequalities \eqref{twoforbidin} hold.
Let $G_1,G_2$ be obtained from $Z(L)$ by replacing the entries $(1,4),(1,6),(4,1),(6,1)$  with the entries given by
 \eqref{defG1G2}, where $\epsilon=1$.
It is straightforward to see that $G_1,G_2\in\cC(3,2)$.  As $Z(L)=\frac{1}{2}(G_1+G_2)$ we deduce that $L$ is nonextreme.

Vice versa, assume that $L$ is nonextreme and $M$ is extreme in $\cM_{1-|y|^2}$.
So $L=\frac{1}{2}(\tilde L_1+\tilde L_2)$ for some $\tilde L_1,\tilde L_2\in\cL_{3,2}$.
Let $G_i=Z(\tilde L_i)$ for $i=1,2$.  Since $M$ is extreme in $\cM_{1-|y|^2}$ it follows that $G_1,G_2$ are obtained from $Z(L)$
by replacing the entries $(1,4),(1,6),(4,1),(6,1)$ with the entries given by \eqref{defG1G2}, where $\epsilon=1$ and $(z,w)\trans\ne \0$.
Apply Lemma \ref{rangelem} to $G_1$ and $G_2$ to deduce \eqref{twoforbidin}.  \qed

\section{Remarks on the additivity conjectures}
Recall that for a density matrix $X\in\rH_{n,+,1}$ the von Neumann entropy $S(X):=-\tr X\log X$ \cite{NC00}.  Clearly, $S(X)\ge 0$ and $S(X)=0$
if and only if $X$ is a pure state, i.e. $\rank X=1$. The minimum entropy output of a quantum channel $L\in\cL_{m,n}$ is defined as
\[S_{\min}(L):=\min_{X\in\rH_{m,+,1}}S(L(X)).\]
Clearly, $S_{\min}(L)\ge 0$ and equality holds if and only if $L(\rH_{m,+,1})$ contains a pure state.  Given two quantum channels $L_i\in\cL_{m_i,n_i}$
for $i=1,2$ it is well known that one can define a tensor product $L_1\otimes L_2$, which is a quantum channel in $\cL_{m_1m_2,n_1n_2}$  \cite{Sho04}.
The minimal characterization of $S_{\min}(\cdot)$ yields
\begin{equation}\label{tenprodin}
S_{\min}(L_1\otimes L_2)\le S_{\min}(L_1)+S_{\min}(L_2).
\end{equation}
The famous \emph{additivity conjecture} claimed equality in the above inequality.  See \cite{Sho04} for several equivalent forms of this conjecture.
It was shown in \cite{GF12} that the minimum entropy output of a quantum channel is locally additive.
Hastings gave a nonconstructive counterexample to the additivity conjecture \cite{Has09}.  A detailed analysis of Hastings' counterexamples show that
it exists in a very high dimension \cite{FKM10}.   So it is of great interest to find counterexamples to the additivity conjecture in small dimensions.

An obvious question is: do the additivity conjectures hold for any pair $L_1,L_2\in\cL_{2,2}$, i.e. qubit channels?
For Holevo capacity the additivity holds if $L_1\in\cL_{2,2}$ is unital, ($L_1(I_2)=I_2$), and $L_2\in \cL_{n,n}$ is arbitrary \cite{Kin02}.

\begin{corol}  Let $L_j\in \cL_{m_j,2,m_j}$ be a quantum channel from qu-$m_j$ to qubit of Choi rank at most $m_j$ for $j=1,2$ . Then
\begin{equation}\label{minoutent0}
S_{\min}(L_1)=S_{\min}(L_2)=S_{\min}(L_1\otimes L_2)=0.
\end{equation}
In particular,  the minimum entropy output is additive in this case.
\end{corol}
\proof
Corollary \ref{condimLmnk} yields that $L_1(\rH_{m_1,+,1}), L_2(\rH_{m_2,+,1})$ contain pure states $Y_1,Y_2$ respectively.  Hence
$(L_1\otimes L_2)(\rH_{m_1m_2,+,1})$ contains
a pure state $Y_1\otimes Y_2$.  Therefore \eqref{minoutent0} holds.  So $S_{\min}(L_1)+S_{\min}(L_2)=S_{\min}(L_1\otimes L_2)$.  \qed

We do not know if the Holevo capacity is additive for two qubit channels if neither of the channels is unital.

 \bibliographystyle{plain}
 
  \end{document}